\documentclass[twocolumn,aps,showpacs,preprintnumbers,amsmath,amssymb,superscriptaddress]{revtex4-1}
\usepackage{graphicx}% Include Fig.~files
\usepackage{dcolumn}% Align table columns on decimal point
\usepackage{bm}% bold math
\usepackage{amsmath}
\usepackage{subfigure}
\usepackage{hyperref}
\usepackage{soul}
\usepackage[normalem]{ulem}
\hypersetup{
    colorlinks,
    citecolor=blue,
    filecolor=black,
    linkcolor=blue,
    urlcolor=blue
}

\begin{document}

%\preprint{APS/123-QED}

\title{Exploring Packaging Strategies of Nano-embedded Thermoelectric Generators}% Force line breaks with \\
%\thanks{A footnote to the article title}%

\author{Aniket Singha}
% \altaffiliation[Also at ]{Physics Department, XYZ University.}%Lines break automatically or can be forced with \\
\affiliation{%
Department of Electrical Engineering,\\
Indian Institute of Technology Bombay, Powai, Mumbai-400076, India\\
 %This line break forced with \textbackslash\textbackslash
}%
\author{Subhendra D. Mahanti}
\affiliation{%
Department of Physics and Astronomy,\\
Michigan State University, East Lansing, MI-48824, USA\\
 %This line break forced with \textbackslash\textbackslash
}%

\author{Bhaskaran Muralidharan}%
 \email{bm@ee.iitb.ac.in}
\affiliation{%
Department of Electrical Engineering,\\
Indian Institute of Technology Bombay, Powai, Mumbai-400076, India\\
 %This line break forced with \textbackslash\textbackslash
}%

%\collaboration{MUSO Collaboration}%\noaffiliation

%\author{Charlie Author}
% \homepage{http://www.Second.institution.edu/~Charlie.Author}
%\affiliation{
 %Second institution and/or address\\
 %This line break forced% with \\
%}%
%\affiliation{
% Third institution, the second for Charlie Author
%}%
%\author{Delta Author}
%\affiliation{%
% Authors' institution and/or address\\
% This line break forced with \textbackslash\textbackslash
%}%

%\collaboration{CLEO Collaboration}%\noaffiliation
\date{\today}% It is always \today, today,
             %  but any date may be explicitly specified

\begin{abstract}
Embedding nanostructures within a bulk matrix is an important practical approach towards the electronic engineering of high performance thermoelectric systems. For power generation applications, it ideally combines the efficiency benefit offered by low dimensional systems along with the high power output advantage offered by bulk systems. In this work, we uncover a few crucial details about how to embed nanowires and nanoflakes in a bulk matrix so that an overall advantage over pure bulk may be achieved. First and foremost, we point out that a performance degradation with respect to bulk is inevitable as the nanostructure transitions to being multi moded. It is then shown that a nano embedded system of suitable cross-section offers a power density advantage over a wide range of efficiencies at higher packing fractions, and this range gradually narrows down to the high efficiency regime, as the packing fraction is reduced. Finally, we introduce a metric - \emph{the advantage factor}, to elucidate quantitatively, the enhancement in the power density offered via nano-embedding at a given efficiency. In the end, we explore the maximum effective width of nano-embedding which serves as a reference in designing  generators in the efficiency range of interest.  
\end{abstract}

\maketitle

%\tableofcontents

\section{Introduction}
Low dimensional systems such as nanowires or nanoflakes are potential candidates for high performance thermoelectric systems with the likely enhancement in their electronic figure of merit \cite{enhance1,enhance2,sny08,Chen_1} and hence the maximum thermodynamic efficiency, which is attributed to a distortion in their electronic density of states \cite{enhance4}. However low-dimensional systems operate at much lower power outputs in comparison to their bulk counterparts. Bulk systems, on the other hand, offer much higher power densities for practical considerations, but at much lower efficiencies. Therefore, it is essential to beat this so called power-efficiency trade-off \cite{bm,reference1} by intelligently designing nano-structured bulk systems \cite{enhance4,sanchez1,akshay} aimed at combining the efficiency benefit of nano structuring with the power density advantage of bulk. 

\begin{figure}[!htb]
\subfigure[]{\includegraphics[scale=0.23]{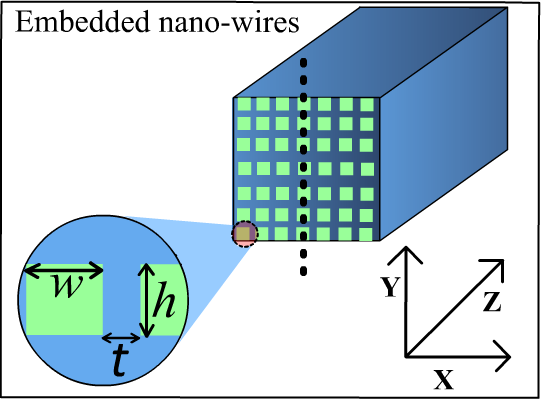}
}\subfigure[]{\includegraphics[scale=0.23]{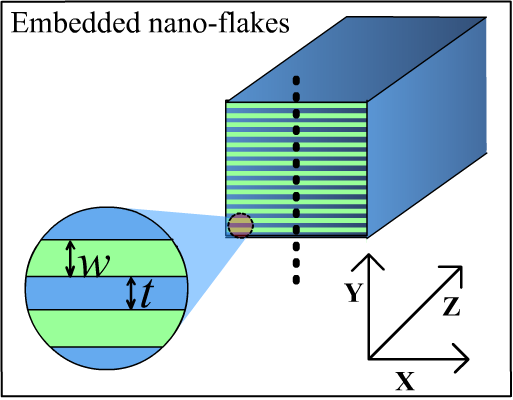}
}
\vspace{.15in}

\subfigure[]{\includegraphics[width=3.4in, height=2.4in]{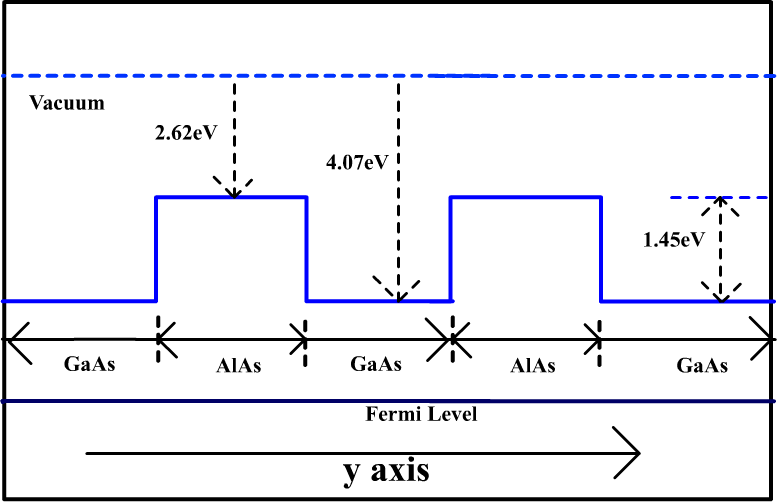}
}
\caption{Device Schematics. (a) Schematic of 1-D quantum wires embedded in a 3-D matrix. (b) Schematic of a 2-D quantum well embedded in a 3-D matrix (c) Band diagram along the $y$ direction depicting a typical GaAs/AlAs structure for the case where GaAs and AlAs have been doped to obtain the same work-function. Band diagrams are shown along the dotted line of (a) and (b)  }
\label{fig:schematic}
\end{figure}

Approaches toward nano-structuring so far have proven successful in the suppression of phonon mediated heat flow \cite{phonon1,phonon2,phonon3,phonon4,phonon5,phonon6,Chen_2}, while electronic engineering based on the DOS \cite{goldsmid,extra1,extra2} of low-dimensional systems \cite{response1,response2,response3,response4} is still a topic of current and active research. One possible approach for electronic engineering is to pack low-dimensional systems in a bulk matrix \cite{reference1,reference2,Chen2} as shown in the schematic in Fig.~\ref{fig:schematic}(a) and (b), such that the efficiency benefit of nano structuring is tapped along the transport direction and the power advantage of bulk is harnessed via the areal packaging strategy \cite{akshay}. With advancements in fabrication and lithographic procedures, atomic layers can be placed with the precision of the order of atomic dimensions \cite{tinytransistor1}, and hence it is envisionable that such packaging strategies are quite possible within the current experimental capabilities. Also, with tremendous progress in oxide based nanoelectronics, it may also be possible to engineer the desired barrier profile on demand \cite{Levy}. 

The key object of this paper is to analyze the packing strategies of such nano-embedded bulk systems and explore under what operating conditions do they offer advantages over traditional bulk thermoelectric generators. Specifically, we consider nanowire and nanoflake-embedded bulk systems, as depicted in Fig.~\ref{fig:schematic}(a) and (b), where the nano structure concerned is packed in between sufficiently thick barriers, as shown in the sample cross-sectional band profile depicted in Fig.~\ref{fig:schematic}(c). An analysis of packing strategies concerns a judicious choice of the barriers between the nano-structures and its relationship with the effective power density at a given operating efficiency to be discussed in depth here.

Traditionally, electronic engineering of low dimensional thermoelectrics concerns the optimization of the electronic figure of merit $z_{el}T$ defined by 
\begin{equation}
z_{el}T=\frac{S^2 \sigma}{\kappa_{el}}T,
\label{eq:zt_def}
\end{equation}
with $S$, $\sigma$ and $\kappa_{el}$ being the Seebeck coefficient, electrical conductivity and electronic thermal conductivity respectively, and $T$ is the operating temperature. 

While low dimensional systems definitely offer the advantage in terms of an enhanced figure of merit, it is now increasingly clear that one has to consider the power-efficiency trade-off \cite{bm,whitney,reference1} for a more complete design strategy. The most glaring example of the power-efficiency trade-off is the Sofo-Mahan thermoelectric \cite{sofo}, which features an infinite electronic figure of merit, but zero power output \cite{bm,whitney,linke}. That being said, $z_{el}T$ is not an accurate measure of efficiency when power output is of the order of the \emph{quantum bound} which is the maximum power any thermoelectric generator can produce  \cite{whitney},  \cite{esposito1},  \cite{esposito2},  \cite{esposito3}. So, in this paper we primarily focus on the power density versus efficiency trade-off in embedded nano-systems.

A few crucial and fundamental aspects of electron engineering via nano-embedding are pointed out in this work. First and foremost, we show that a performance degradation is inevitable as the nano structure transitions to being multi moded, a trend which has been noted in a few experiments \cite{phonon6}. Turning to packing strategies, we point out that at higher packing fractions, judicious nano-embedding offers a significant power density advantage when operated over a large range of efficiencies. However, at lower packing fractions, this range gradually narrows down to the high efficiency regime. Finally, we introduce a metric - \emph{the advantage factor}, to elucidate quantitatively, the enhancement in the power density offered via nano-embedding at a given efficiency. At the end, we explore the maximum effective width of nano-embedding which serves as a reference in designing nano-embedded generators in the efficiency range of interest. 
\begin{figure}[!htb]
\hspace{-.6cm}\subfigure[]{\includegraphics[scale=0.175]{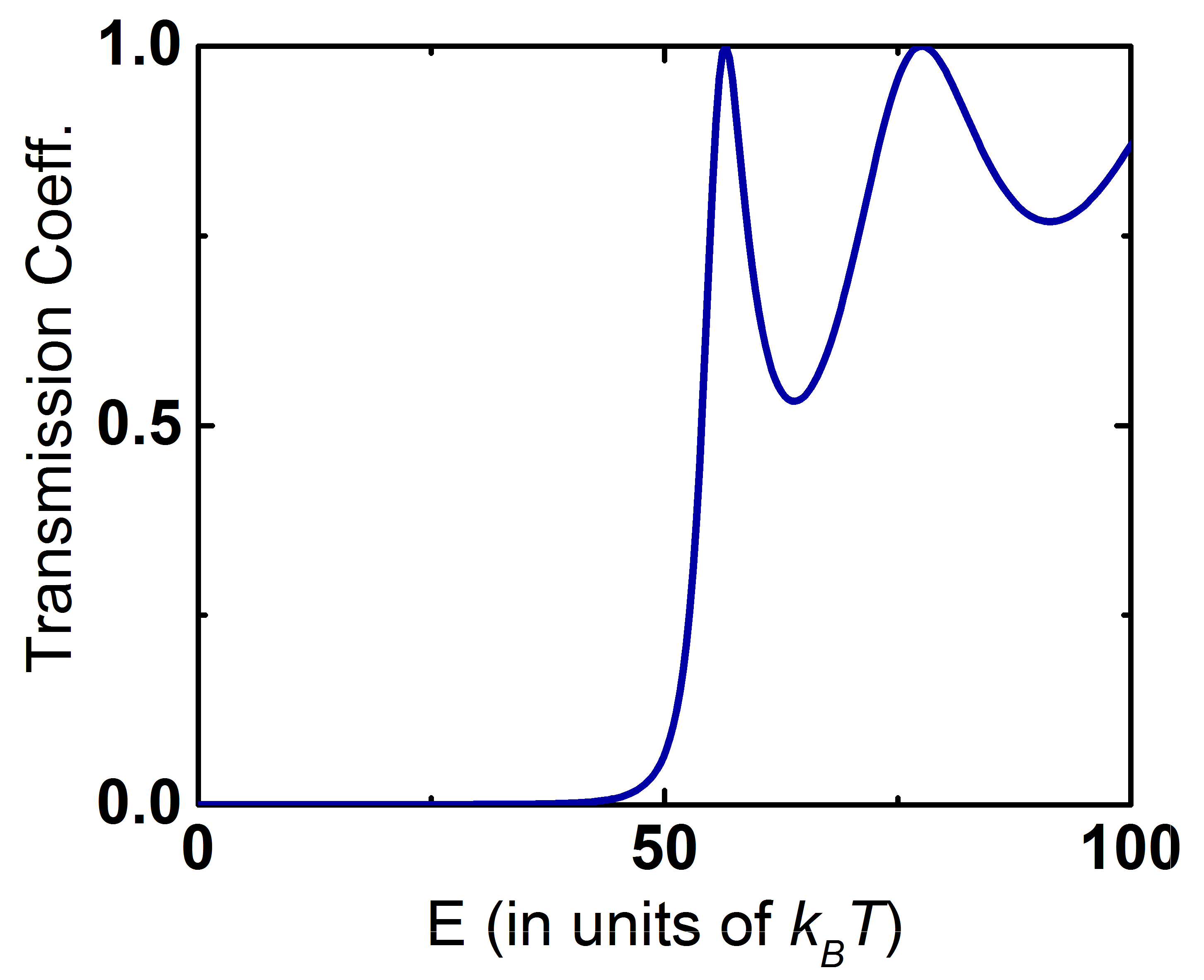}

}\subfigure[]{\includegraphics[scale=0.18]{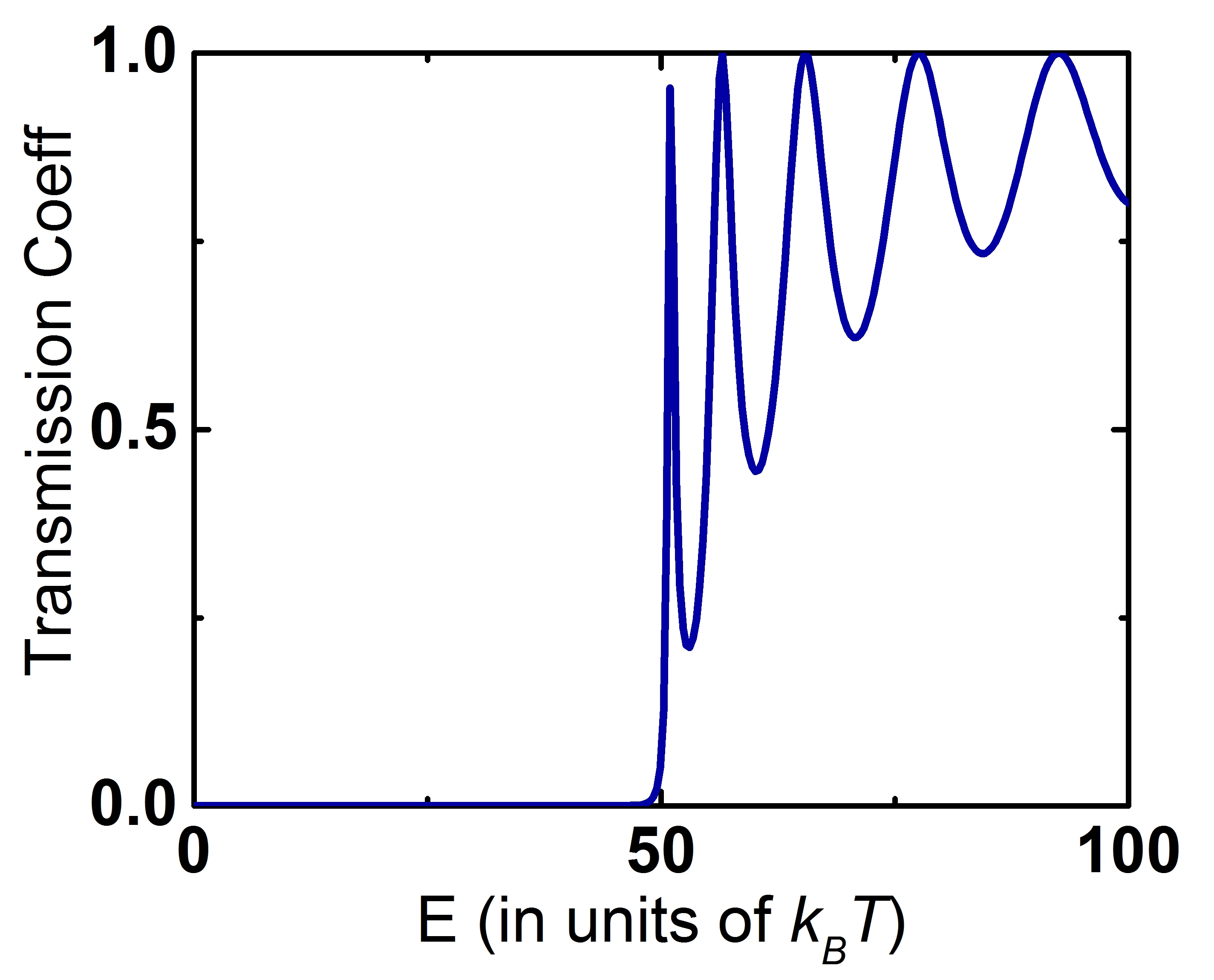}

}

 \protect\caption{Engineering optimum packing separation between nano-structures. Transmission coefficient along the $y$ direction through (a) a $5nm$ AlAs energy barrier (b) $10nm$ AlAs barrier of barrier height 1.45eV as shown in Fig.~\ref{fig:schematic}(c). It can be noted from (a) and (b) that the $10 nm$ barrier offers not more significant advantage in comparison with the $5nm$ barrier.}
\label{fig:transmissioncoeff}

\end{figure}

\section{Model Details}
We model nano-embedded systems as quantum wires/flakes separated by energy barriers as shown in Fig.~\ref{fig:schematic}(a) and (b). For the purpose of our calculation, we  use  the parameters of lightly doped GaAs conducting channels and AlAs potential barriers to separate two GaAs conducting channels. AlAs serves as a potential candidate for energy barrier fabrication because 
GaAs and AlAs have  approximately same lattice constant resulting in minimization of lattice strain due to lattice constant mismatch  \cite{streetman}.
We start our calculations with  nano devices
of cross-sectional dimension $2.82nm$ embedded in a 3-D bulk matrix.  The electron affinity of GaAs and AlAs are $4.07eV$ and  $2.62eV$ respectively. According to Anderson's rule \cite{streetman}, an energy barrier of $1.45eV$  in a GaAs/AlAs system is feasible if the two materials are doped in a way such that their work-functions are equal. Further manipulation of the doping concentration of these materials results in higher energy barriers. The conduction band minima of the GaAs/AlAs system, given equal work-functions, are schematically depicted in Fig.~\ref{fig:schematic}(c). For our discussions we will simply assume that Anderson's rule is valid and neglect complicated realistic variations in band offsets. A more practical DFT based calculation of band offsets in GaAs/AlAs junctions is depicted in \cite{offset1, offset2}.

The ability of a 5nm AlAs barrier to suppress electronic transmission between two adjacent GaAs conducting channels is depicted in Fig.~\ref{fig:transmissioncoeff}(a). The vanishing values of the transmission coefficient along the $y$ direction over a specific energy range implies an absence of electronic coupling between adjacent GaAs channels in that energy range. The transmission coefficient is calculated in ballistic limit using the non-equilibrium Green's function (NEGF) formalism with an atomistic tight-binding Hamiltonian \cite{datta,akshay}. It is clearly seen from Fig.~\ref{fig:transmissioncoeff}(a) that a $5nm$ AlAs energy barrier of $1.45eV$ can suppress electronic transmission between adjacent GaAs channels upto $E_k=50k_BT$ at $T=330K$ where $E_k$ is the kinetic energy of electrons. Fig.~\ref{fig:transmissioncoeff}(b) demonstrates that  a $10nm$ AlAs energy barrier is only marginally advantageous over a $5nm$ AlAs barrier in inhibiting electronic coupling below a kinetic energy of $E_k=50k_BT$.  We therefore conclude that $5nm$ is an optimum separation between two adjacent GaAs conducting channels in a GaAs/AlAs system.

%To calculate the transmission function through the devices,
%we employ Non-Equilibrium Green's Function formalism. For charge and
%heat current calculation, the transmission function is used in Landauer
%equations.

We consider ballistic thermoelectric generators with electron transport in the $z$ direction such that $T(E_z)\thickapprox1$ \cite{datta,akshay}. The charge and heat current through a device is given by  \cite{datta}
\begin{equation}
\begin{split}
I_{C}=\frac{2q}{h}\underset{s}{\sum}\int T(E_z)[f_{H}(E_z+E_s) \\
-f_{C}(E_z+E_s)]dE_Z
\end{split}
\label{eqn:chargecurrent}
\end{equation}

\begin{equation}
\begin{split}
I_{Q}=\frac{2}{h}\underset{s}{\sum}\int(E_z+E_s-\mu_{H})T(E_z) 
[f_{H}(E_z+E_s) \\ 
-f_{C}(E_z+E_s)]dE_z,
\end{split}
\label{eqn:heatcurrent} 
\end{equation}
where $I_C$ is the charge current, $I_Q$ is the heat current, $E_z$ is the kinetic energy along the transport direction, $E_s$ is the minimum kinetic energy of $s^{th}$ sub-band, $f_{H/C}$ is the electron occupancy probability at the hot/cold contacts given by the respective Fermi-Dirac distributions kept at electrochemical potentials $\mu_{H/C}$. The summation in \eqref{eqn:chargecurrent} and \eqref{eqn:heatcurrent} extends over all conducting sub-bands.

In the subsequent calculations, the temperatures of the hot and cold contacts, labeled $H$ and $C$, are assumed to be $330K$ and $300K$ respectively. For simplicity, we assume parabolic dispersion $E-k$ relationship with a constant
effective mass $m^{*}=0.07m_e$. 
To make calculations simple, we assume that both the contacts are
perfectly symmetric and the doping of the GaAs conducting channel is uniform and 
that the voltage drops linearly across the GaAs channels. Assuming
that the contacts are in local equilibrium, the quasi-Fermi levels
in the contacts is given by $\mu_{H/C}=\mu_{0}\pm V/2$ where $V$ is
the voltage dropped across the thermoelectric generator due to current flow through an external circuit
and $\mu_{0}$ is the equilibrium electrochemical potential across the entire device.

To generate the power density vs efficiency plots in the next section, we use a voltage controlled model \cite{bm,akshay} in order to vary the bias voltage across the thermoelectric generator. The power density and efficiency ($\eta$) are calculated for each value of the bias voltage ($V$). The set of points comprising the power density and efficiency then constitute the curves in the $\eta-P$ plane for a particular $\mu_0$.

With the above description in mind, we now turn our attention to the power density-efficiency trade-off in three types of embedded nano systems in the absence of electron-phonon interaction (ballistic limit)
and phonon mediated heat flow- (1) A single moded nanowire of cross-section
$2.82nm\times2.82nm$ (2) A 2-D nanoflake of width $2.82nm$ (3) A multi moded nanowire ($28.2nm\times 28.2nm$) packed in a matrix. 

\section{Nanostructures versus Bulk}
{\it{Single moded nanowires:}} Our calculations use a single moded nanowire of length less than the mean free path of electrons
so that electronic transport occurs predominantly in the ballistic
regime. For simplicity, we assume that the electron wavefunction is
confined only within the GaAs conducting channels. The kinetic energy of the electrons
with respect to the conduction band minimum is given by 
\begin{equation}
E=E_{s}(x,y)+\frac{\hslash^{2}k_{z}^{2}}{2m^{*}},
\end{equation}
where $E_{s}(x,y)$ is the minimum  energy of the $s^{th}$ sub-band and $m^{*}=0.07m_{e}$ is the conduction band effective mass of the electrons in GaAs. 

A nanowire exhibits single moded property when the thermal broadening of the Fermi-Dirac distribution function $k_BT$ is much less than the minimum energy separation between two sub-bands. Assuming that the separation
between two sub-bands must be greater than $5k_BT$, we deduce that the maximum cross-sectional dimensions that preserves the single moded property of a GaAs nanowire is approximately $11.2nm$. We also note that the maximum cross-sectional dimension depends on the effective
mass of the electrons. For example, for 
$m^{*}=m_{e}$, redoing the above calculation gives $max\{w,h\}\leq2.96nm$.
\begin{figure}[!htb]
\subfigure[]{\includegraphics[scale=0.17]{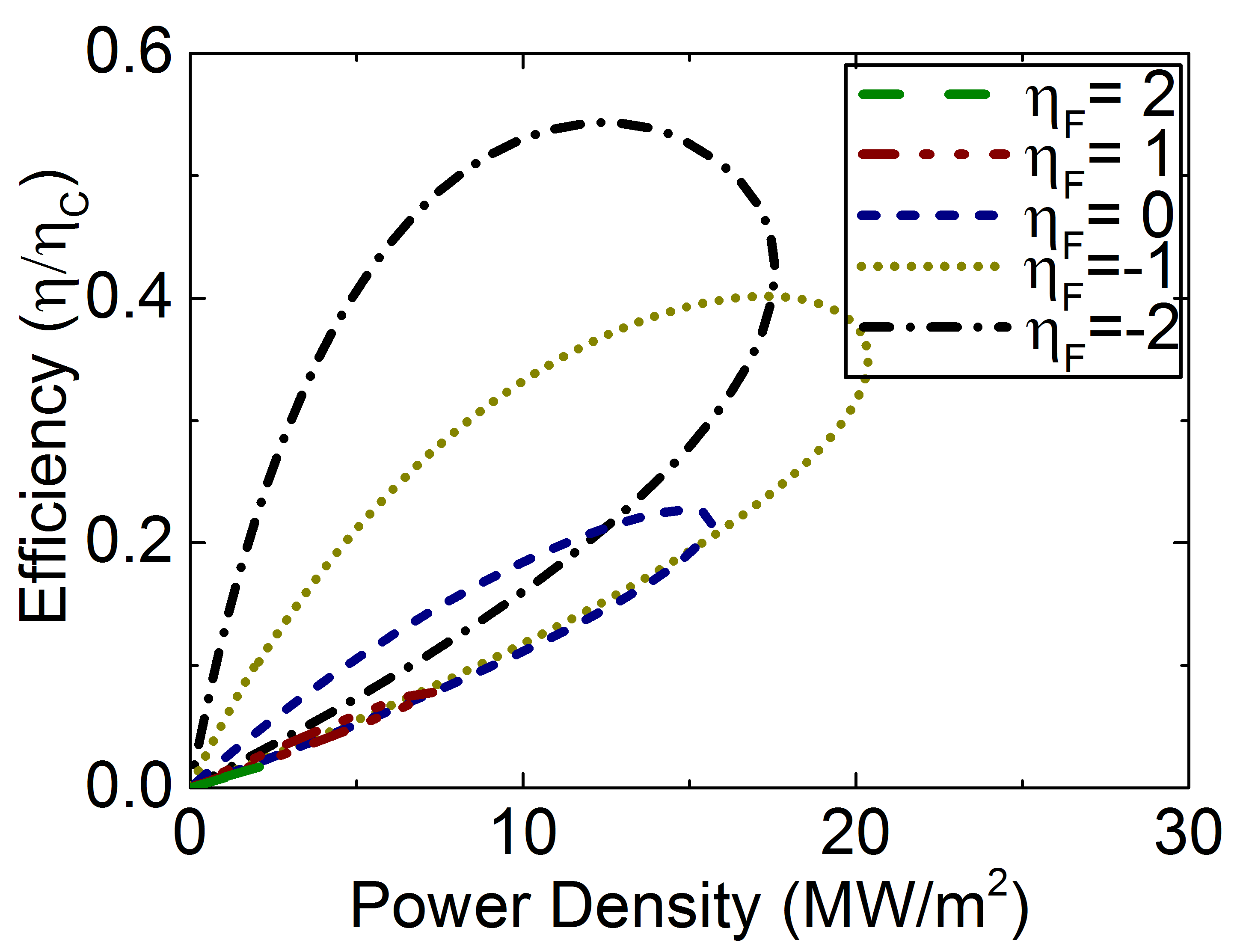}

}\subfigure[]{\includegraphics[scale=0.17]{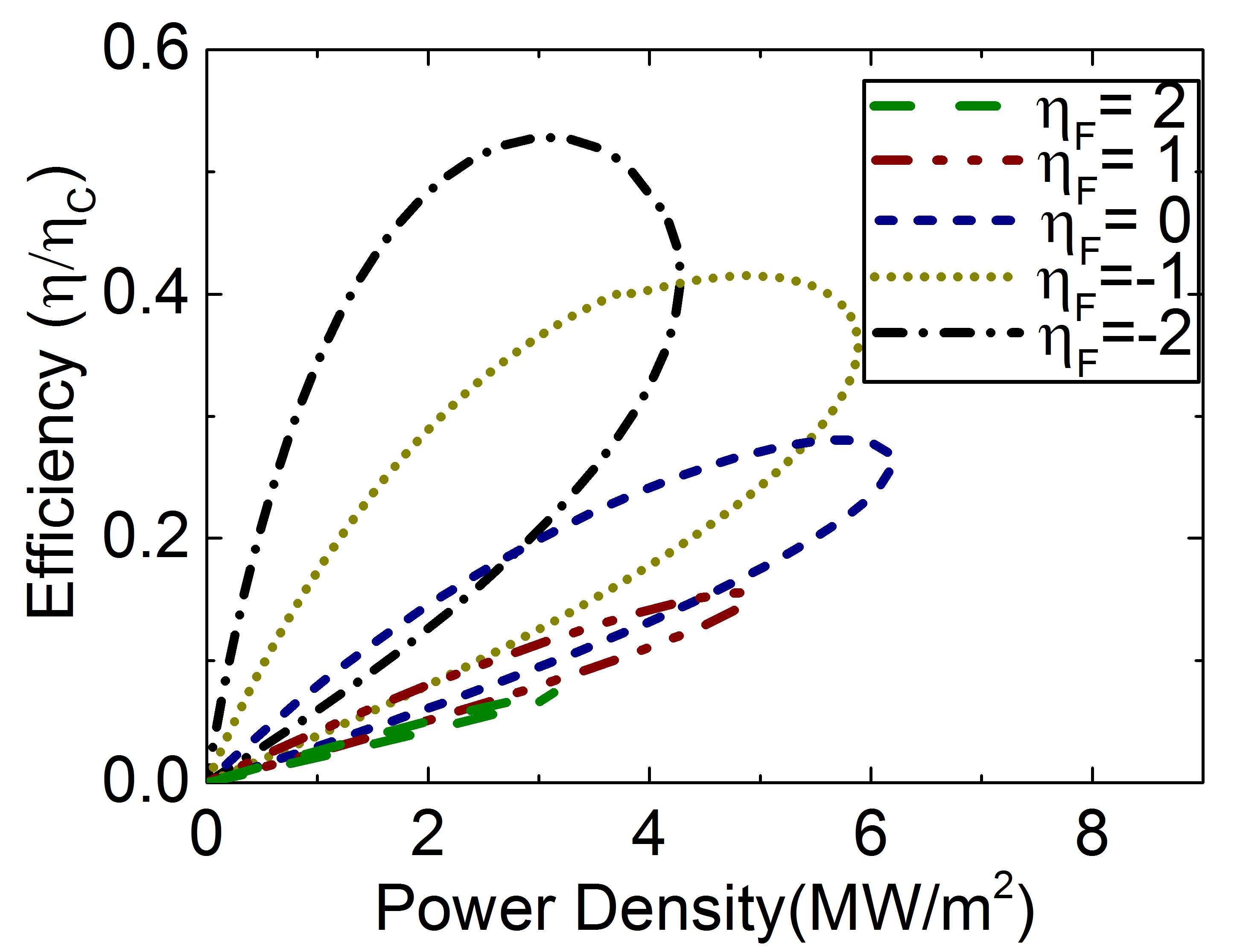}

}

\subfigure[]{\includegraphics[scale=0.17]{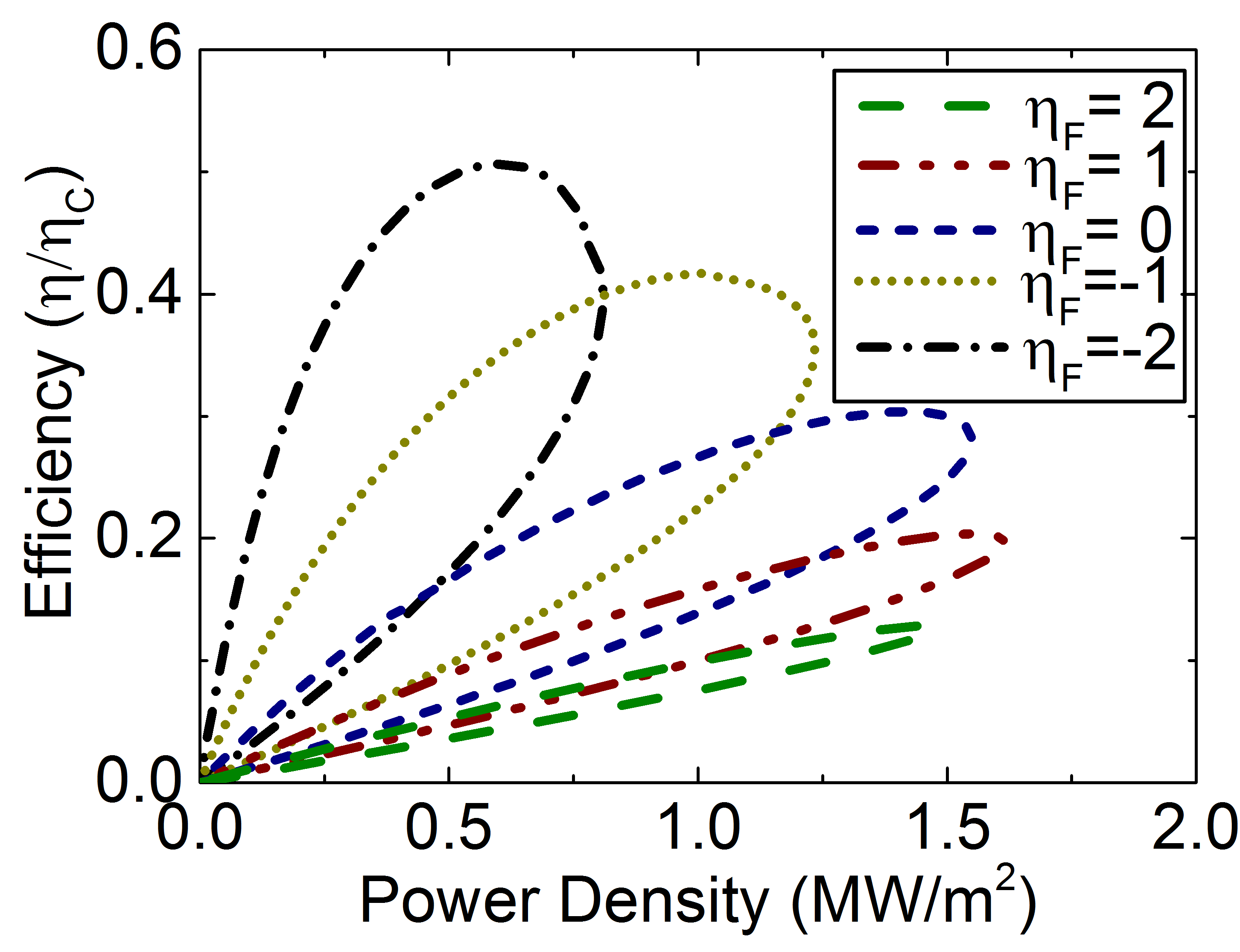}

}\subfigure[]{\includegraphics[scale=0.17]{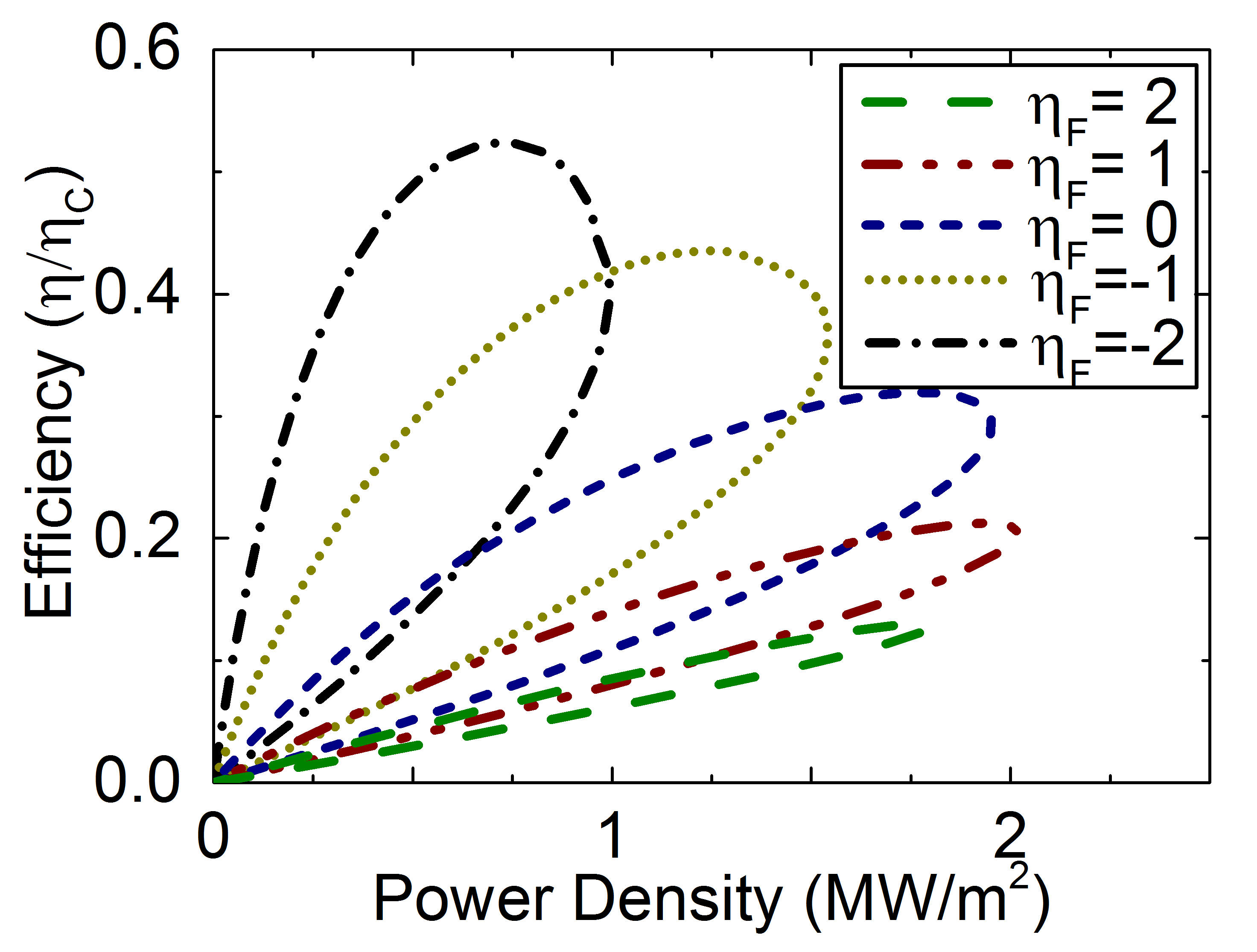}}

\protect\caption{Power-efficiency trade-off. Power Density vs efficiency of a (a) $2.82nm$x$2.82nm$ single moded quantum wire (b) $2.82nm$ quantum flake (c) $28.2nm$x$28.2nm$ multi moded nano-scale wire and (d) 3-D bulk structure. Plots are shown for values of $\eta_F$ ranging from $-2$ to $+2$ with $\eta_F=\frac{E_s-\mu_0}{k_BT}$, where, $E_s$ is the lowest energy of the first sub-band.}
\label{fig:powerefficiency}
\end{figure}

Although our calculations show that a nanowire remains single moded for $max(h,w)<11.2nm$, the true potential of nanowire generators is harnessed at $max(h,w)<<11.2nm$. This is because the total power generated by the single moded nanowire does not change with area making the power density inversely proportional to the cross-sectional area. The power efficiency trade-off for a $2.82nm\times 2.82nm$ nanowire is shown in Fig. \ref{fig:powerefficiency}(a). For comparison, power-efficiency plot of a bulk thermoelectric generator is also shown in Fig.~\ref{fig:powerefficiency}(d). We note that the maximum power density of a  single moded quantum wire of cross-sectional dimensions $2.82nm\times 2.82nm$ is approximately $20MW/m^2$, while that of a bulk generator is of the order of $2MW/m^2$. Such high power density is expected given that the maximum power produced by a single moded nanowire is of the order of $0.17nW$ \cite{reference2}. \\
{\it{Nanoflakes:}} As in the previous case, we consider a nanoflake of length
less than the mean free path of the electrons such that electronic transport is ballistic. Assuming that the transport is in the $z$ direction and the electronic confinement is in the $y$ direction, the kinetic energy of the itinerant electrons is given by: 
\begin{equation}
E=E_{s}(y)+\frac{\hslash^{2}(k_{z}^{2}+k_{x}^{2})}{2m^{*}},
\end{equation}
where $E_{s}(y)$ is the minimum  energy of the $s^{th}$ sub-band due to electronic confinement and $m^{*}=0.07m_{e}$
is the effective mass of the conduction band electrons in GaAs. 

Purely two-dimensional electronic band properties are imposed on the nanoflakes when the thermal broadening of the Fermi function ($k_BT$) is much less than the minimum energy separation between two sub-bands. Assuming that the separation between two sub-bands must be greater than $5k_BT$ at, we get
$w\leq11.2nm$.

Similar to a single moded nanowire, the power output of a nanoflake with only one conducting sub-band  does not change with the width of the nanoflake. In terms of power density, a nanoflake with the smallest possible width is the most advantageous. Fig.~\ref{fig:powerefficiency}(b) shows the power-density versus efficiency plot of a $2.82nm$ wide nanoflake. The maximum power density of a nanoflake of width $2.82nm$ is around $6.1MW/m^2$, while a bulk generator provides a maximum power density of $2MW/m^2$. \\
{\it{Multi moded nanowires:}} Next, we consider a nanowire with discrete sub-bands such
that energy separation between the lower modes is of the order of thermal broadening of the Fermi-Dirac distribution function ($k_BT$).  Fig.~\ref{fig:powerefficiency}(c) shows the power-efficiency trade-off in this case. It can be seen in comparing Figs.~\ref{fig:powerefficiency}(a), (b), (c) and (d) that a nanowire
with multiple modes provides a lower performance not only
in power density but also in efficiency  compared to the single moded nanowire, nanoflake and bulk thermoelectric generator. 
 
A single moded nanowire operates at a higher efficiency at maximum power density compared to a multi moded nanowire. Multiple conducting
sub-bands provide a  number of higher energy levels for electron transport. Electronic
flow through these higher energy levels from the hot contact to the cold
contact results in higher electronic heat conductivity thereby reducing
the efficiency of the thermoelectric generator. Also positioning  the Fermi level
near the higher energy sub-bands may result in significant back flow of electrons
from the cold contact to the hot contact through the lower energy sub-band (hole like conduction) thereby
reducing the power output and the voltage obtained at a given efficiency.

By comparing Figs.~\ref{fig:powerefficiency}(a),(b),(c) and (d), we can infer that single moded nanowires and nanoflakes offer the possibility for a power density advantage, and hence may be engineered for an enhancement in the performance via nano-embedding as opposed to multi moded nanowires. So, we conclude that there is a maximum cross-sectional dimension beyond which the performance of a nanowire may degrade below the bulk performance, which is what we will consider next.\\
\begin{figure}[!htb]
\subfigure[]{\includegraphics[scale=0.085]{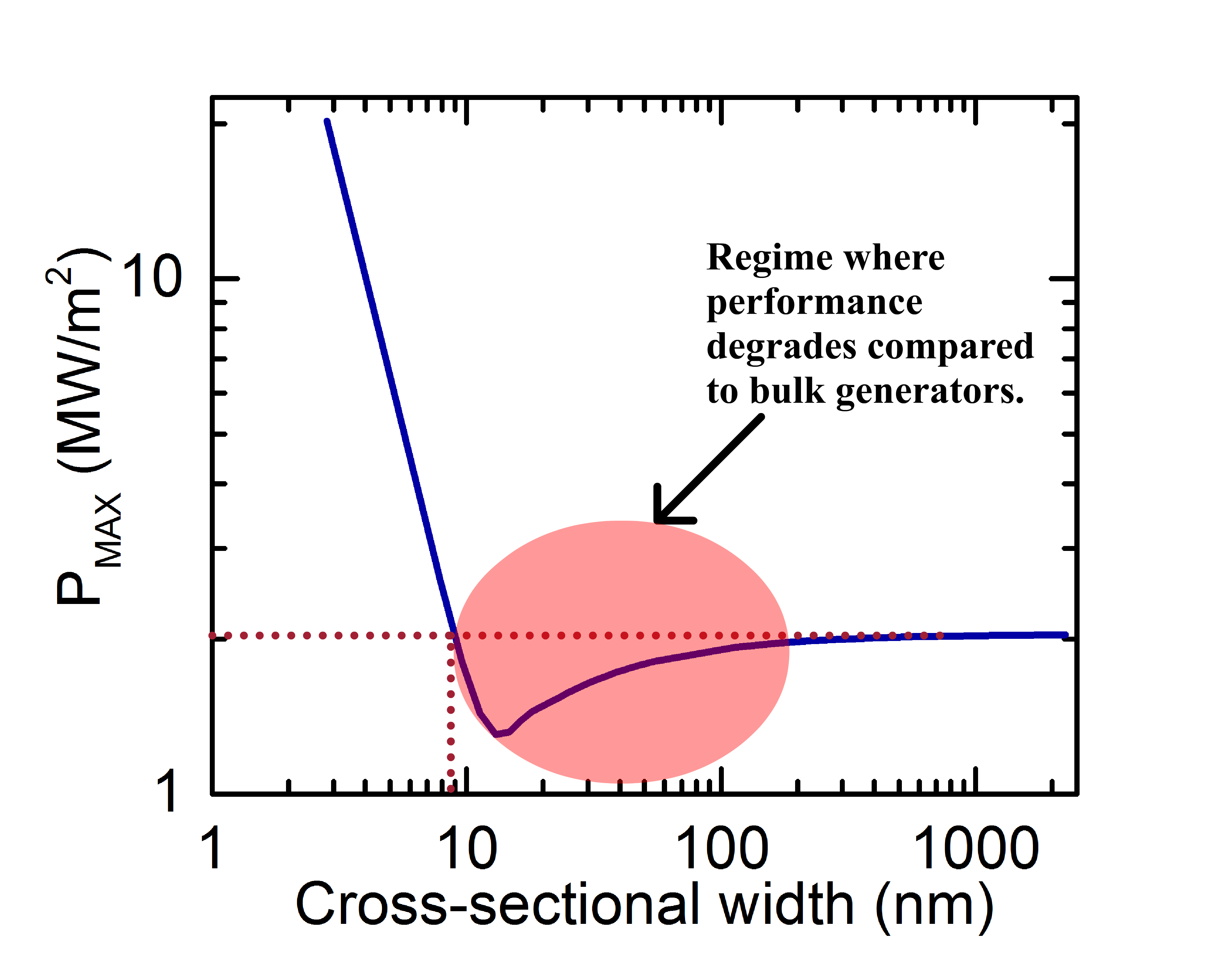}

}

\hspace{-.2cm}\subfigure[]{\includegraphics[scale=0.085]{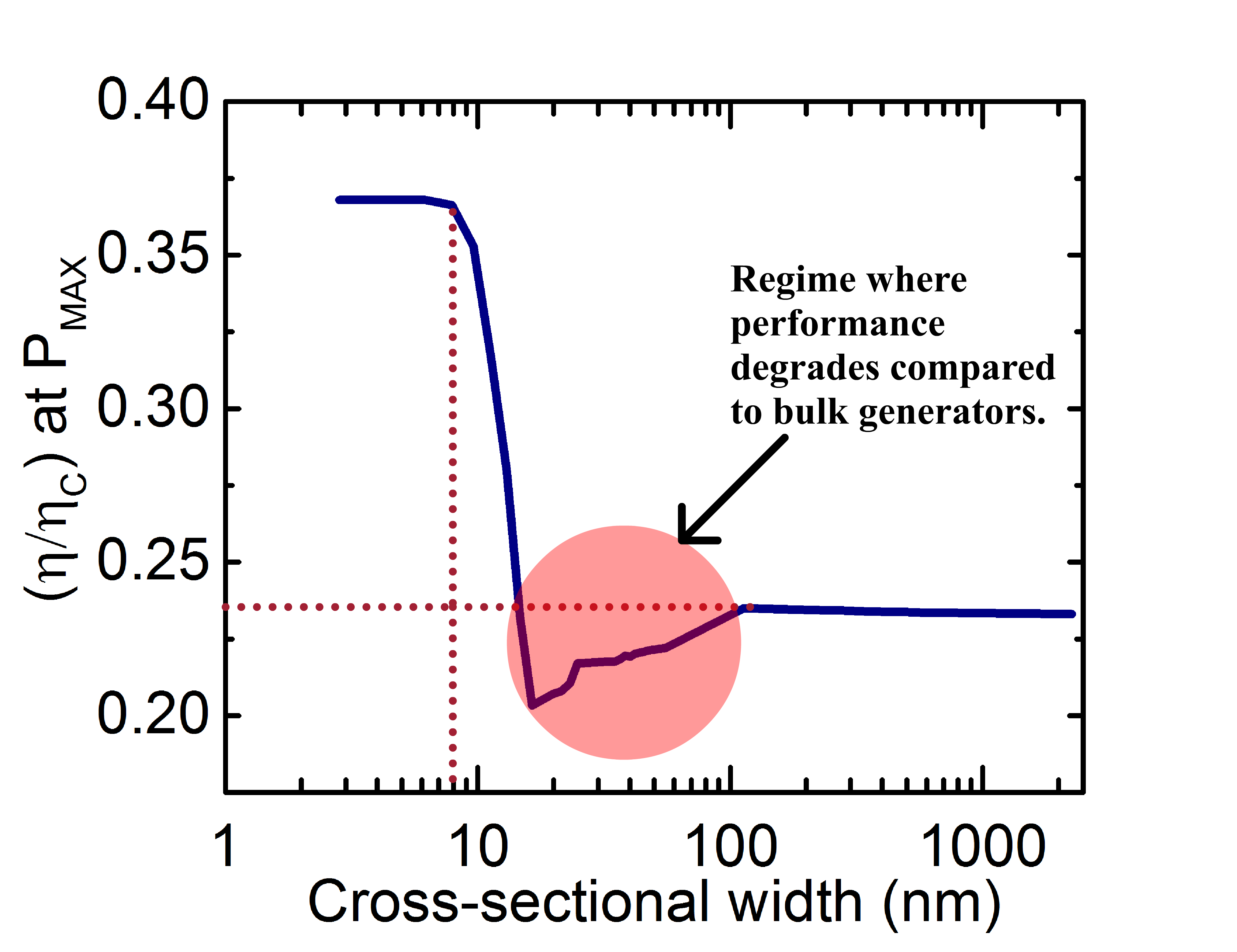}}

\protect\caption{Transition from nano-scale to bulk (a) in the case of a nanowire of square cross-section in terms of the maximum power generated. (b) Change in efficiency at maximum power as a quantum wire makes transition from being single moded to bulk. The region marked in pink depicts the region where there is no utility in using nanowires for power generation applications.  }
\label{fig:transition}

\end{figure}
{\it{Single moded nanowire to bulk transition:}} The maximum cross-sectional width for which nanowires retain single moded electronic properties from the perspective of power generation is of particular interest in designing thermoelectric generators. For large cross-sectional dimension of the nanowire in comparison with the lattice constant, the separation between two sub-bands $\Delta E$ becomes much
less than $k_{B}T$, and the nanowire begins to show bulk properties. Two quantities of interest which can be used to distinguish between lower dimensional and bulk thermoelectric generators are maximum power density ($P_{MAX}$) and the efficiency at maximum power density ($\eta _{_{P_{MAX}}}$). 

In Figs.~\ref{fig:transition}(a) and (b), we show that the advantages of single moded nanowires may be harnessed for $max(h,w)\leq8nm$, where the logarithm of the power density decreases linearly and the efficiency at maximum power ($\eta_{P_{MAX}}$) remains almost constant with the logarithm of the cross-sectional dimensions. Between $8nm\leq (h,w)\leq200nm$, it behaves as a multi moded nanowire where the difference $\Delta E$  between two sub-bands is of the order of $k_BT$. For $l\geq200nm$ the nanowire behaves as a bulk material depicting the expected trend of constant power-density. As stated previously, the performance of a  nanowire in the multi moded regime degrades in terms of both the maximum power density as seen in Fig.~\ref{fig:transition}(a) as well as the efficiency as seen in  Fig.~\ref{fig:transition}(b) in comparison to single moded nanowire and bulk thermoelectric generators. A multi moded nanowire hence shows no promise in applications to nano-embedded systems. 

\begin{figure}[!htb]
	\centering
		\includegraphics[width=0.50\textwidth]{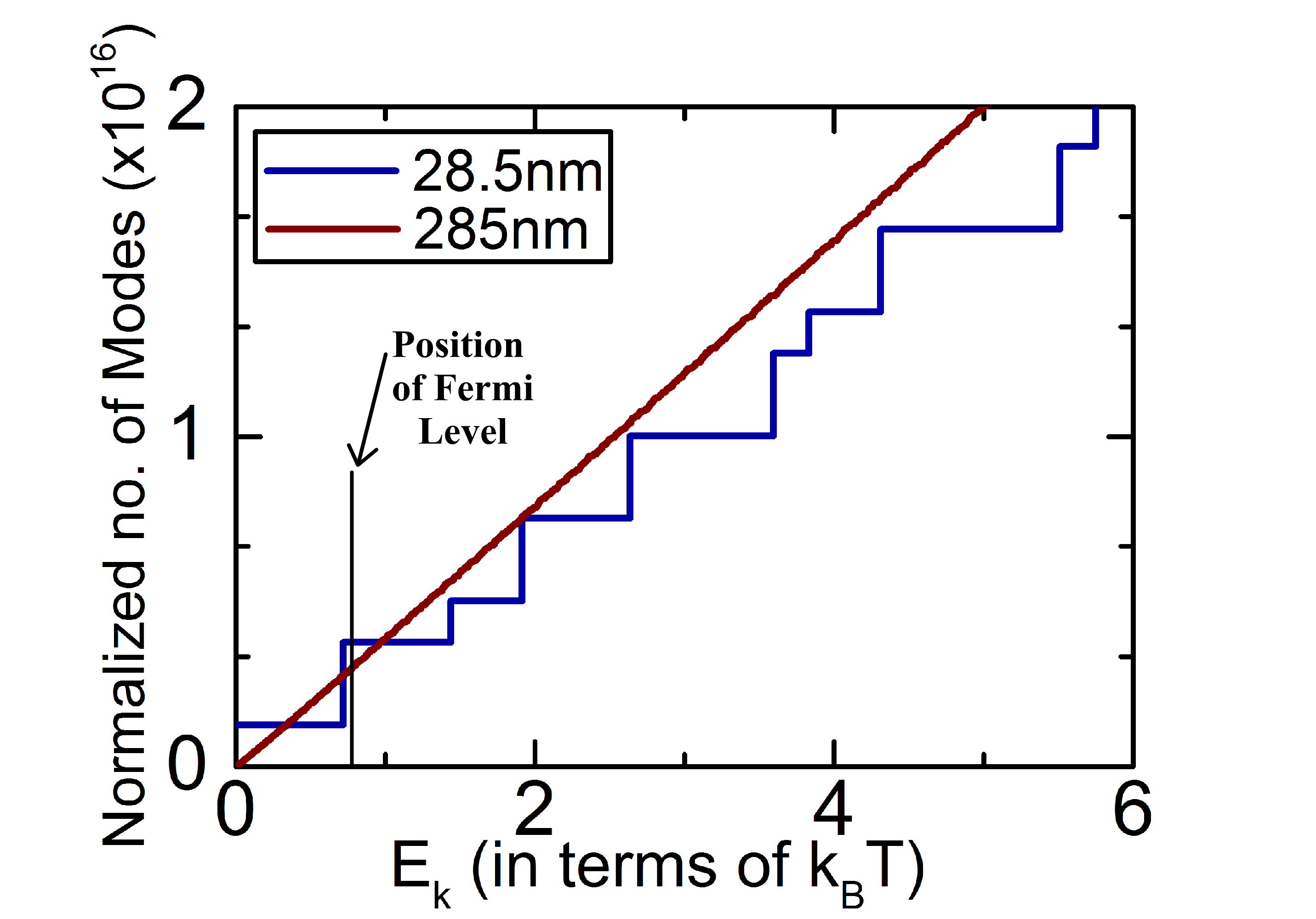}
	\caption{Normalized modes per unit area profile as a function of energy from the minimum of the first sub-band. Above the Fermi energy, the normalized number of modes of a 282nm square nanowire is either greater than or equal to that for the 28.2nm nanowire. Greater number of modes facilitate higher current density against the applied bias resulting in an increased power density in the case of the 282nm nanowire compared to that of the 28.2nm nanowire.}
	\label{fig:modes}
\end{figure}

The reason for the degradation of power density in multi moded nanowires compared to bulk is evident from Fig.~\ref{fig:modes}, which shows the normalized mode density per unit area of a 28.2nm  and a 282nm wide square nanowire. Above the Fermi energy, the 282nm nanowire contains a higher number of modes per unit area which results in higher output power density. The degradation of the efficiency at maximum power is not so evident via a similar line of argument, since the efficiency is a ratio between the power and the heat current \cite{bm,akshay}. It is therefore of some use to make a back of the envelope calculation by considering $z_{el}T$ for the two structures. In the linear response regime, the efficiency at maximum power is given by \cite{maxp}
\begin{equation}
\eta\thickapprox\frac{\eta_C}{2}\frac{z_{el}T}{z_{el}T+2},
\end{equation}
which is a monotonically increasing function of $z_{el}T$. The calculated value of $z_{el}T$ for the $28.2nm$ and $282nm$ wide nanowires are $1.4828$ and $1.6672$ respectively, which proves that a nanowire of width $282nm$ would operate at higher $\eta_{P_{MAX}}$ compared to a nanowire of width $28.2nm$. The $z_{el}T$ calculations performed above are through circuit arguments \cite{bm,akshay}, which match the calculations performed by evaluating all the linear response coefficients in \eqref{eq:zt_def}.

The domain of each regime in Figs.~\ref{fig:transition}(a) and (b), however, would scale inversely as the effective mass ($m^*$) of the material and the temperature of operation $T$. Thus, nanowires made of materials with higher effective mass would require smaller cross-section to harness the advantages of being single moded. 

\section{Operating Regimes of Nano-embedded Generators}
We will now consider the effect of packing nanostructures and the details involved in the process of nano-embedding the structures considered so far. 
Embedded nanowires of cross-sectional dimensions $h$ and $w$, with a separation $t$ between adjacent nanowires, provide a packing fraction of $\frac{wh}{wh+(w+h)t+t^{2}}$. Similarly embedded nanoflakes of width $w$ separated by energy barriers of width $t$ provide a packing fraction of $\frac{w}{w+t}$. The details of the calculations are carried out in the appendix. It is of foremost importance to note that despite the higher power density provided by a single moded nanowire of cross-section $2.82nm\times 2.82nm$, its effective power density in a matrix is limited due to a very low packing fraction.

A matrix packed with nanowires of square cross-section ($2.82nm\times 2.82nm$) separated by a $5nm$ energy barrier provides a packing fraction of 0.13. Similarly, $2.82nm$ nanoflakes packed in a matrix provide a packing fraction of 0.361, when adjacent flakes are separated by a $5nm$ energy barrier. These numbers are reduced to $0.0485$ and $0.2203$ respectively, if the separation between successive nanowires/flakes is increased to $10nm$. We now turn our attention to the operating points of the respective thermoelectric generators taking their packing fraction into account. We use the term \emph{effective power density} to denote the power per unit area of a matrix packed with nanowires/flakes. 

\begin{figure}[!htb]
\subfigure[]{\includegraphics[scale=0.085]{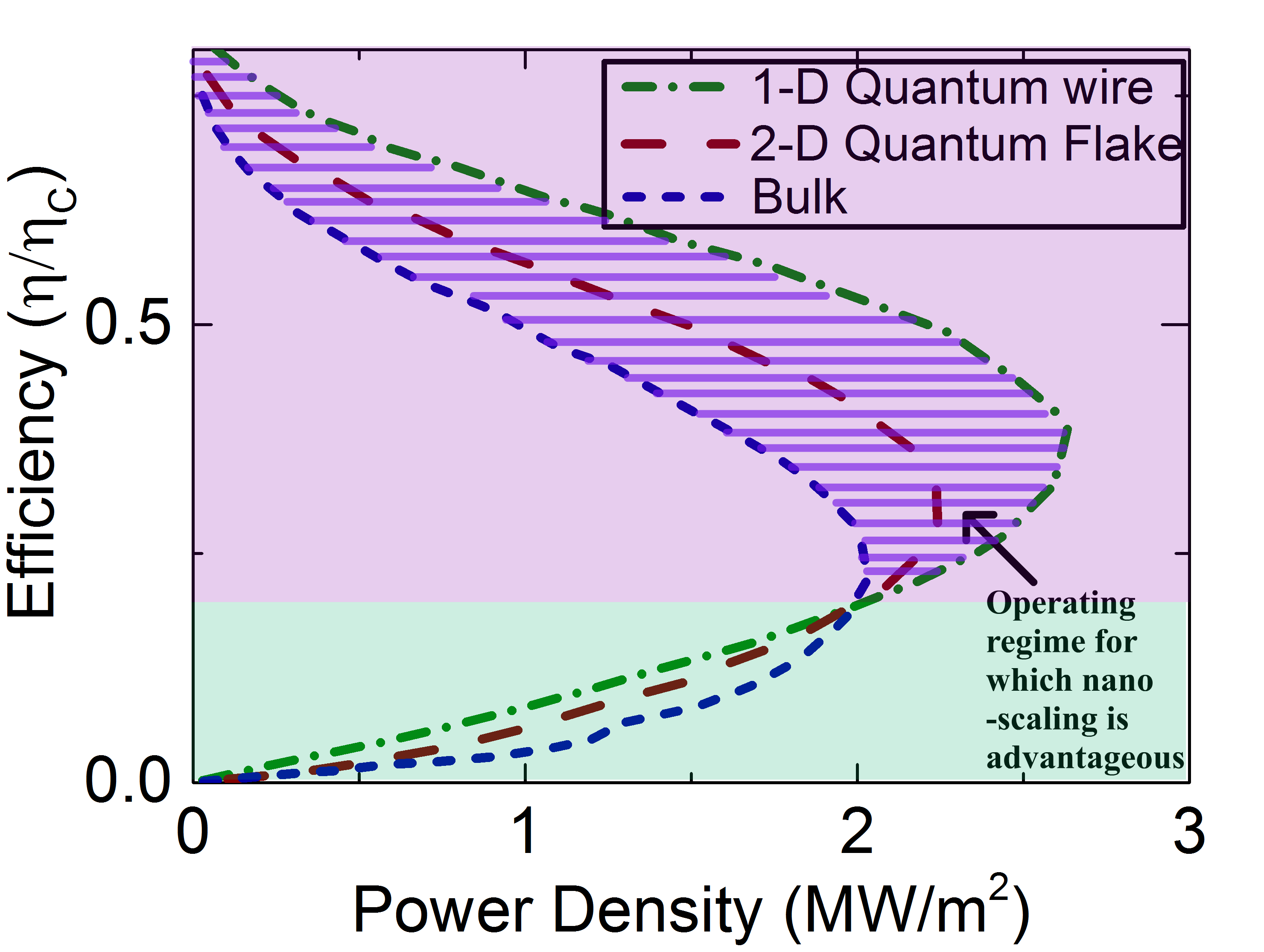}

}

\subfigure[]{\includegraphics[scale=0.085]{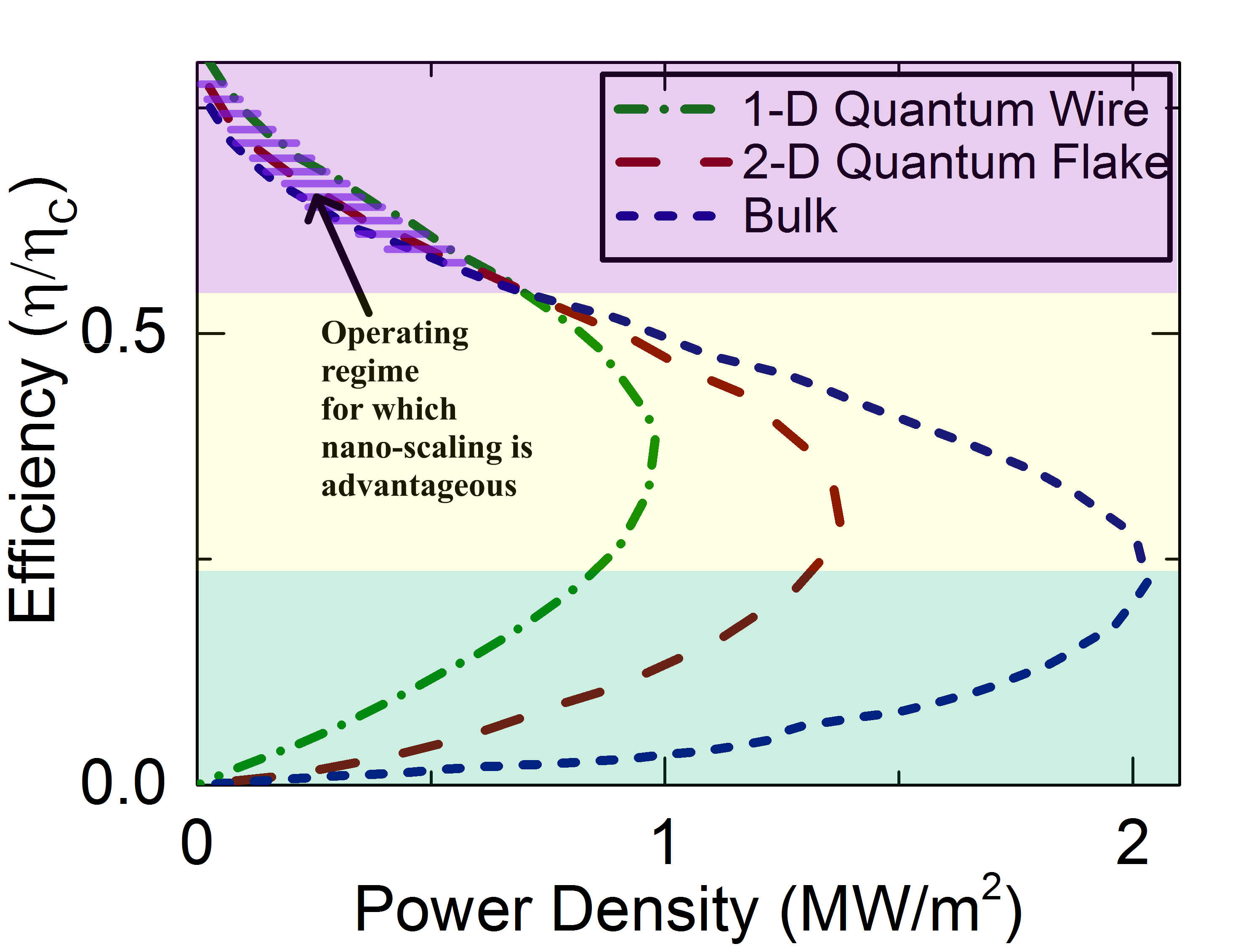}

}

\protect\caption{Operating Regions. Efficiency vs. effective power density plots for a single moded quantum wire of cross-sectional dimension $2.82nm\times 2.82nm$, a 2-D quantum flake of width $2.82nm$, and 3-D bulk taking packing fraction into account for (a) $5nm$ separation barrier between two nano-structures (b) $10nm$ separation barrier between two nano structure. It can be seen in comparing (a) and (b) that nano-embedding offers a significant advantage at lower packing fractions over a wide range of efficiencies. }
\label{fig:operatingpoints}

\end{figure}

In the context of our discussion, operating points are defined as points
in the effective power density-efficiency plane ($\eta-P$) where the effective power-density is maximum for a given efficiency or the efficiency is maximum for a given effective power-density.
For power generation applications, these points offer the minimum effective power density-efficiency tradeoff. For bulk or nano-embedded generators, depending on the cross-sectional dimensions, material parameters and packing fractions, there is a set of points in the $\eta -P$ plane where the device can operate \cite{whitney}, which we term as \emph{allowed points}. The subset of all these points constitute the allowed region of operation in the  $\eta -P$ plane. All other points in the $\eta -P$ plane constitute the \emph{forbidden region }\cite{whitney}. Operating points lie on the boundary of the allowed region and the forbidden region of operation. These are the points along which the maximum effective power density at a given efficiency, or equivalently, maximum efficiency at a given effective power density, can be obtained. Fig.~\ref{fig:operatingpoints} show the operating
points of embedded nano-systems for a separation of $5nm$ and $10nm$ respectively. Each figure can be seen to be divided into three regimes:

{\it{Regime 1:}} In this regime, there is no trade-off between power and
efficiency for the bulk thermoelectric generators, that is, efficiency
increases monotonically with increasing power. Thermoelectric generators are not operated in this regime. This regime is marked using the \emph{light blue} colour in Figs.~\ref{fig:operatingpoints}(a) and (b).

{\it{Regime 2}}: In this regime, there is a trade-off between power and efficiency
for the bulk generators and bulk generators perform better in terms of the power density in comparison to nano-embedded systems. Using the bulk generator is advantageous in this regime. This regime is marked using the \emph{light yellow} color in Figs.~\ref{fig:operatingpoints}(a) and (b).

{\it{Regime 3}}: In this regime, there is a trade-off between power and efficiency
for the bulk generator, but nano embedded systems perform better than the bulk in
terms of power density at a given efficiency. In this region of operation, nano-embedding
is advantageous. This regime is marked using the \emph{light purple} color in Fig.~\ref{fig:operatingpoints}(a) and (b).

It is shown in Fig.~\ref{fig:operatingpoints}(a) that for nanowires/flake embedded systems of cross-sectional
dimensions $2.82nm$ separated by $5nm$ energy barrier, the second regime does not exist. That is, for all relevant points of operation, nano-embedding is advantageous. However when the separation between adjacent nanowires/flakes is increased to $10nm$, as noted in Fig.~\ref{fig:operatingpoints}(b), nano-embedding is advantageous
only in the regime of higher efficiency. This could be  understood
by the fact that increasing separation between adjacent nanowires/flakes 
decreases the packing density resulting in a decrease in the effective power output per unit area. 

\begin{figure}[!]
\hspace{-.2in}\includegraphics[scale=0.36]{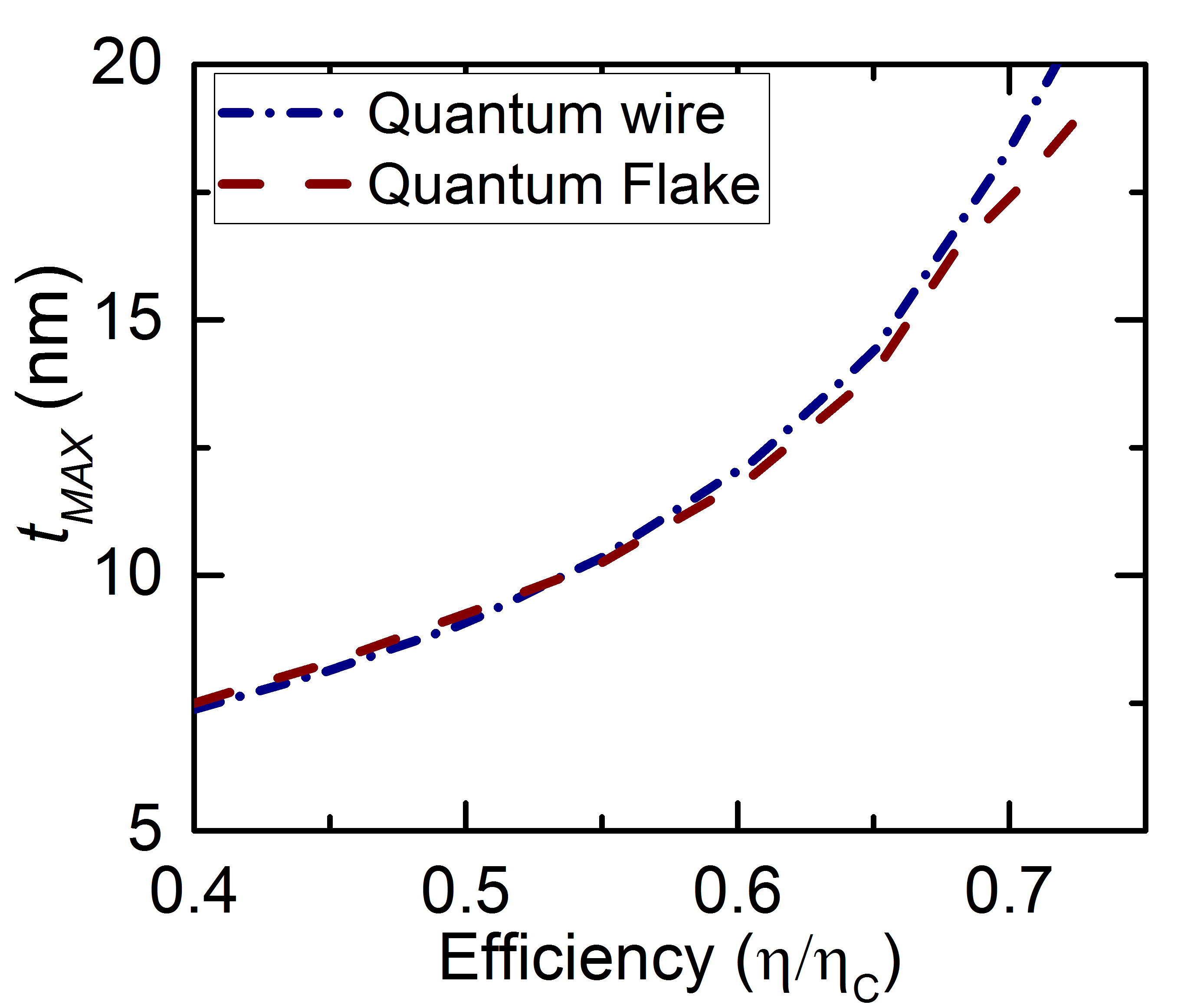}

\protect\caption{Maximum width of the energy barrier that can be used for a single moded nanowire of cross sectional dimensions $2.82nm\times 2.82nm$ and 2-D quantum flake of width of $2.82nm$. For a separation width greater than the given the maximum values, bulk provides a better effective power density-efficiency tradeoff. }
\label{fig:tunwidth}

\end{figure}

In this context, it may be of particular interest to know what is the
maximum width by which two nano devices may be separated
while still retaining the power-efficiency advantage out of nano-embedding. This is shown in Fig.~\ref{fig:tunwidth}, where the maximum possible separation
between two nano-scale devices increases when the device is operated
at a higher efficiency. In other words, for the same separation 
between two nano-generators, nano-scaling 
offers a higher power advantage when they are operated in higher efficiency
regime.

\section{Advantage Factor due to Nano-Scaling} \label{advantagefactor}

A metric to quantify the power advantage offered by nano-embedded systems over their bulk counterpart serves as a deciding factor for nano-scaling thermoelectric generators. The parameter of interest, is indeed, the factor by which the power density is increased in nano embedded systems over bulk generators.

\begin{figure}[!htb]

\subfigure[]{\includegraphics[scale=0.33]{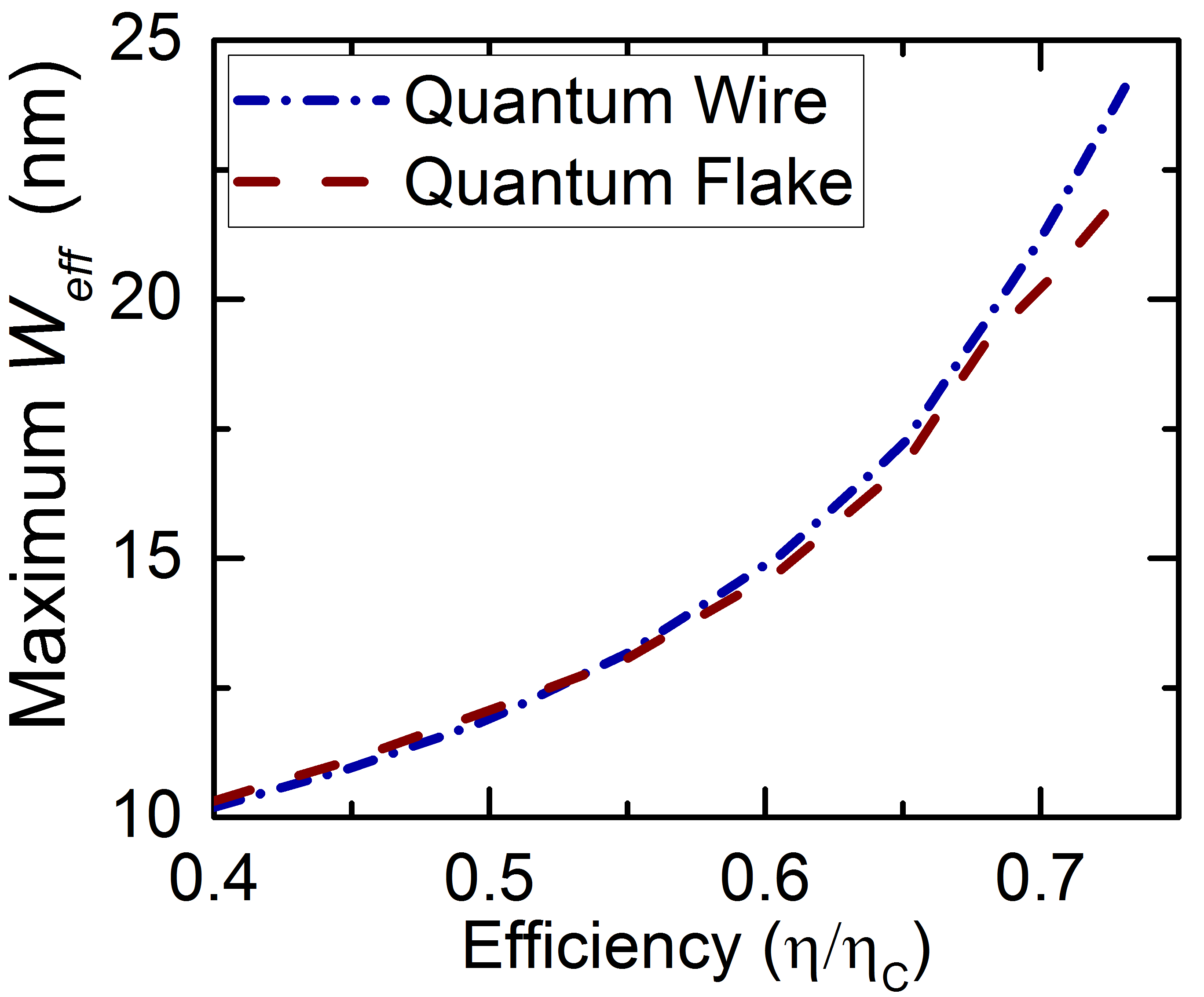}
}

\subfigure[]{\hspace{.8cm}\includegraphics[scale=0.31]{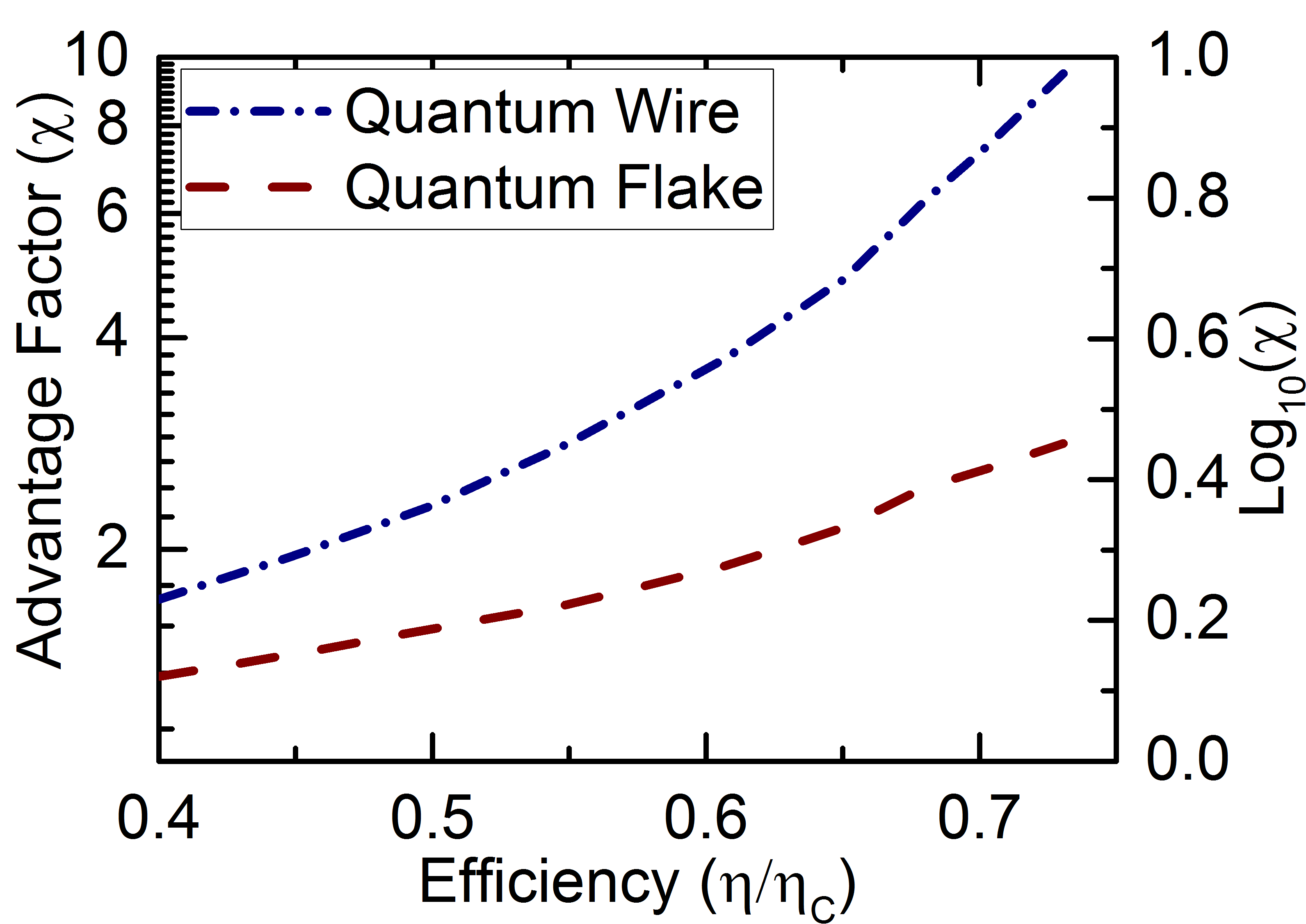}
}

\protect\caption{(a)Maximum \emph{effective width} ($W_{eff}$)  of  square quantum wires and  quantum flakes embedded in a bulk matrix. The maximum value $W_{eff}$ depends on the effective mass of the material which is used to describe nano-scale thermoelectric generators. (b)Advantage factor for square cross-sectional quantum wires and quantum flakes of cross-sectional dimension $2.82nm$ separated by $5nm$ energy barriers. }
%(Refer to section \ref{advantagefactor}) 
\label{fig:advfac}

\end{figure}

In the context of our discussion, we define an \emph{advantage factor} at a given efficiency $\eta$ as
\[ 
\chi(\eta)=\frac{P_{nano}(\eta)}{P_{bulk}(\eta)},
\]
where $P_{nano}$ is the effective power density of a nano-embedded thermoelectric generator and $P_{bulk}$ is the power density of the bulk thermoelectric generator. The advantage factor is  a basic measure of the advantage we get from nano-embedding.  For a given set of material parameters, the advantage factor depends on (a) the efficiency of operation of the thermoelectric generators and (b) the \emph{effective area} per nano-generator. We define effective area of a nano-generator as 
\[
A_{eff}=\frac{1}{N},
\]
where $N$ is the number of nano generators embedded in a square matrix of unit cross-sectional area. Fig.~\ref{fig:advfac}(b)  shows the logarithm of the advantage factor versus the efficiency for  nano-embedded GaAs generators of cross-sectional dimension $2.82nm$ separated by a $5nm$ AlAs energy barrier.  Doubling the effective area per unit nano-scale generator decreases $log(\chi)$ at each point on Fig.~\ref{fig:advfac} by approximately 0.3dB.  For square embedded nanowires of cross-sectional dimensions \emph{w} separated by energy barriers of width \emph{t}, the \emph{effective area} is given by $A_{eff}=(w+t)^2$.

For embedded nanoflakes of width \emph{w} and separated by energy barriers of width \emph{t}, the \emph{effective area} is given by $
A_{eff}=w+t$. The quantity $(w+t)$ is therefore a parameter of importance and we call it the \emph{effective width} ($W_{eff}$) of the nano-embedded generators.  A lower $W_{eff}$ results in a higher advantage factor. Fig.~\ref{fig:advfac}(a) shows the maximum effective width that can be allocated to a unit nano embedded generator with the effective width maximum when $\chi =1$. In terms of the maximum effective width, the advantage factor may be written as: $\chi =(\frac{(W_{eff})_{MAX}}{W_{eff}})^2$, for a single moded nanowire and $\chi =\frac{(W_{eff})_{MAX}}{W_{eff}}$ for a nanoflake with a single mode along confined direction. For the same effective width, nanowires perform better than nanoflakes, the difference in their advantage factors being an increasing function of the efficiency. It should be noted that advantage factor is calculated along the operating points at a given efficiency for single moded nanowires and flakes. 

\section{Conclusion}

In this work, we have pointed out a few crucial aspects to be considered for the {\it{electronic engineering}} of nano-embedded bulk thermoelectric generators. First and foremost, we pointed out that a performance degradation is inevitable as the nanostructure transitions to being multi moded, a trend which has been noted in a few experiments \cite{majum, heath}. Turning to packing strategies, we pointed out that at higher packing fractions, nano-embedding offers a significant power density advantage when operated over a large range of efficiencies. However, at lower packing fractions, nano-embedding performs poorly, displaying only a marginal advantage and that too, only when operated at much higher efficiencies. Finally, we introduced a metric - \emph{the advantage factor}, to elucidate quantitatively, the enhancement in the power density offered via nano-embedding at a given efficiency. At the end, we explored the maximum effective width of nano-embedding which serves as a reference in designing nano-embedded generators in the efficiency range of interest. From our detailed analysis, we conclude that although  a nanowire/flake is capable of demonstrating enhanced nano-scale thermoelectric properties \cite{enhance1,enhance2} upto a maximum cross-sectional dimensions of $8nm$, it is advantageous to fix the width of nanowires/flakes to the smallest possible value permitted by current fabrication capabilities. In all our calculations, we have neglected decrease in efficiency due to lattice heat conductivity. Although the performance of both nano-embedded generators and traditional generators are degraded due to  heat conduction via phonons, nano-structuring offers a predominant advantage over traditional generators in terms of reduced lattice conductivity. The high density of interfaces in nano embedded generators on the length scale of phonon mean free path effectively scatters long wavelength phonons thereby reducing the lattice conductivity \cite{phonon1,phonon2,phonon3,phonon4,phonon5,phonon6,Chen_2,phonon7} and counteracting the reduction in efficiency to a large extent. Analysis and results obtained in this work should provide general design guidelines to embed nanostructures within a bulk matrix to offer a greater advantage over traditional bulk thermoelectrics, within the limits of current fabrication technologies.

\section*{Appendix}

\subsection{Evaluation of Packing Fraction}

Let us assume rectangular nano-structures embedded in a bulk matrix.
Let the height and width of each nano-scale device be $h$ and $w$
respectively. Assuming the cross-sectional dimensions of the bulk
material (say $H$ and $W$) to be much greater than the cross-sectional
dimensions of the nano-structures and the minimum separation width
between two nano-scale structures to be $t,$ the total number of
nano structures that can be embedded in the bulk matrix is 
\[
N=\frac{HW}{(h+t)(w+t)}.
\]
In the context of our discussion, the packing fraction is defined as the
fraction of the 3-D matrix which is occupied by the embedded nano structures.
From this definition, the packing fraction for embedded nano structures in
the matrix is

\[
P.Fr=Nhw/HW=\frac{\frac{HW}{(h+t)(w+t)}hw}{HW}=\frac{hw}{(h+t)(w+t)}.
\]
For square 1-D nano-scale structures, $h=w$. The packing fraction
then becomes

\[
P.Fr=(\frac{w}{w+t})^{2}.
\]

For 2-D nano-scale structures, we have $h\rightarrow\infty$. So,
the packing fraction is given by 
\[
PFr=Lt_{h\rightarrow\infty}|\frac{hw}{hw+(h+w)t+t^{2}}|=\frac{w}{w+t}
\]

\subsection{Calculation of maximum \emph{effective width} ($W_{eff}$)}

Let us assume the power density of bulk thermoelectric generator is $P_{bulk}$ at
an efficiency $\eta$. At the same efficiency, let power output of a unit embedded nanowire be $P_{1-D}$. We calculate the  effective width of unit nano generator when the overall power density of the bulk is equal to
the overall power density of the embedded nano generators. The number of single moded nanowires required to match the
power density of the bulk is:

\[
N_{min}=\frac{P_{bulk}}{P_{1-D}}.
\]

We need to pack a minimum $N$$_{min}$ number of single moded nanowires per unit area of the matrix to achieve the power density similar to a bulk generator. To gain the advantage embedded nanowires, $N$ nanowires must be packed per unit area of the  matrix such that

\[
N\geq N_{min}.
\]

Assuming the nanowires to be of square cross-section, the maximum effective area which can be allocated for a unit nano generator is :

\[
(A_{eff})_{MAX}=\frac{1}{N_{min}}\geq\frac{1}{N}=A_{eff}
\]

So, the maximum effective width of unit nano generator should be

\[
W_{eff}\leq(W_{eff})_{MAX}=\sqrt{(A_{eff})_{MAX}}
\]
\[
=\frac{1}{\sqrt{N_{min}}}=\sqrt{\frac{P_{1-D}}{P_{bulk}}}.
\]

Similar derivations for 2-D nano-scale devices lead to:
\[
W_{eff}\leq\frac{P_{2-D}}{P_{bulk}},
\]
where $P_{2D}$ is the power output of a 2-D quantum flake which has only one conducting mode along the confined dimension.

% The \nocite command causes all entries in a bibliography to be printed out
% whether or not they are actually referenced in the text. This is appropriate
% for the sample file to show the different styles of references, but authors
% most likely will not want to use it.
%\nocite{*}

\bibliography{apssamp}% Produces the bibliography via BibTeX.

\end{document}